%% file: main.tex
\pdfoutput=1
\documentclass[12pt,a4paper]{article}

\usepackage{ifthen} 
\newboolean{pdflatex}
\setboolean{pdflatex}{true} 

\newboolean{articletitles}
\setboolean{articletitles}{true} 

\newboolean{uprightparticles}
\setboolean{uprightparticles}{false} 

\def\paperauthors{LHCb collaboration} 
\def\paperasciititle{Strong constraints on the b->sgamma photon polarisation from B0->K*0e+e- decays} 
\def\papertitle{Strong constraints\\ on the $b \to s\gamma$ photon polarisation\\ from $\Bd\to\Kstarz e^+ e^-$ decays} 
\def\paperkeywords{{High Energy Physics}, {LHCb}, {flavour physics}, {B physics}, {polarisation}, {rare decay}} 
\def\papercopyright{\the\year\ CERN for the benefit of the LHCb collaboration} 
\def\paperlicence{CC BY 4.0 licence}
\def\paperlicenceurl{https://creativecommons.org/licenses/by/4.0/}

\input{preamble}
\usepackage{longtable} 

\def\meeKstar {\ensuremath{ m(K^+\pi^- \epem)}}

\def\BdToeeKst    {\decay{\Bd}{\Kstarz\ep\en}}
\def\BsToPhiee    {\decay{\Bs}{\phi\ep\en}}
\def\LbToeeL    {\decay{\Lb}{\Pp K^- \ep\en}}
\def\BdToJPsieeKst  {\decay{\Bd}{\jpsi(\epem)\Kstarz}}
\def\ATD     {\ensuremath{A_{\mathrm{T}}^{(2)}}\xspace}
\def\ATIm     {\ensuremath{A_{\mathrm{T}}^{\mathrm{Im}}}\xspace}
\def\ATRe     {\ensuremath{A_{\mathrm{T}}^{\mathrm{Re}}}\xspace}
\def\FL       {\ensuremath{F_{\mathrm{L}}}\xspace}

\def\ctl       {\ensuremath{\cos{\theta_\ell}}\xspace}
\def\ctk       {\ensuremath{\cos{\theta_K}}\xspace}
\def\phit       {\ensuremath{\tilde\phi}\xspace}

\begin{document}

\renewcommand{\thefootnote}{\fnsymbol{footnote}}
\setcounter{footnote}{1}

\input{title-LHCb-PAPER}

\renewcommand{\thefootnote}{\arabic{footnote}}
\setcounter{footnote}{0}

\cleardoublepage


\pagestyle{plain} 
\setcounter{page}{1}
\pagenumbering{arabic}


\input{body}

\input{acknowledgements}




\addcontentsline{toc}{section}{References}
\bibliographystyle{LHCb}
\bibliography{main,standard,LHCb-PAPER,LHCb-CONF,LHCb-DP,LHCb-TDR}

\newpage
\input{LHCb_Authorship_07-Jul-2020}

\end{document}

%% file: preamble.tex
\usepackage[top=1in, bottom=1.25in, left=1in, right=1in]{geometry}

\usepackage{microtype}
\usepackage{lineno}  
\usepackage{xspace} 
\usepackage{caption} 

\usepackage{graphicx}  
\usepackage{color}
\usepackage{colortbl}
\graphicspath{{./figs/}} 

\usepackage{amsmath} 
\usepackage{amssymb}
\usepackage{amsfonts}
\usepackage{upgreek} 

\newcommand*\patchAmsMathEnvironmentForLineno[1]{%
\expandafter\let\csname old#1\expandafter\endcsname\csname #1\endcsname
\expandafter\let\csname oldend#1\expandafter\endcsname\csname
end#1\endcsname
 \renewenvironment{#1}%
   {\linenomath\csname old#1\endcsname}%
   {\csname oldend#1\endcsname\endlinenomath}%
}
\newcommand*\patchBothAmsMathEnvironmentsForLineno[1]{%
  \patchAmsMathEnvironmentForLineno{#1}%
  \patchAmsMathEnvironmentForLineno{#1*}%
}
\AtBeginDocument{%
\patchBothAmsMathEnvironmentsForLineno{equation}%
\patchBothAmsMathEnvironmentsForLineno{align}%
\patchBothAmsMathEnvironmentsForLineno{flalign}%
\patchBothAmsMathEnvironmentsForLineno{alignat}%
\patchBothAmsMathEnvironmentsForLineno{gather}%
\patchBothAmsMathEnvironmentsForLineno{multline}%
\patchBothAmsMathEnvironmentsForLineno{eqnarray}%
}

\usepackage{hyperxmp}

\usepackage[pdftex,
            pdfauthor={\paperauthors},
            pdftitle={\paperasciititle},
            pdfkeywords={\paperkeywords},
            pdfcopyright={Copyright (C) \papercopyright},
            pdflicenseurl={\paperlicenceurl}]{hyperref}

\usepackage[colorinlistoftodos,textsize=scriptsize]{todonotes}

\usepackage[bottom,flushmargin,hang,multiple]{footmisc}

\usepackage[all]{hypcap} 

\input{lhcb-symbols-def} 
\def\flavio     {\mbox{\textsc{Flavio}}\xspace}

\usepackage{cite} 
\usepackage{mciteplus}

%% file: lhcb-symbols-def.tex
\usepackage{xspace} 
\usepackage{upgreek}

\def\lhcb   {\mbox{LHCb}\xspace}

\def\babar  {\mbox{BaBar}\xspace}
\def\belle  {\mbox{Belle}\xspace}

\def\rich   {RICH\xspace}

\def\ecal   {ECAL\xspace}

\def\MagUp {\mbox{\em Mag\kern -0.05em Up}\xspace}

\ifthenelse{\boolean{uprightparticles}}%
{
 
 \def\Pgamma      {\ensuremath{\upgamma}\xspace}

 \def\Pmu         {\ensuremath{\upmu}\xspace}

 \def\Ppi         {\ensuremath{\uppi}\xspace}

 \def\Ppsi        {\ensuremath{\uppsi}\xspace}

 \def\PDelta      {\ensuremath{\Delta}\xspace}                 
 \def\PXi         {\ensuremath{\Xi}\xspace}                 
 \def\PLambda     {\ensuremath{\Lambda}\xspace}                 
 \def\PSigma      {\ensuremath{\Sigma}\xspace}                 
 \def\POmega      {\ensuremath{\Omega}\xspace}                 
 \def\PUpsilon    {\ensuremath{\Upsilon}\xspace}

 \def\PB      {\ensuremath{\mathrm{B}}\xspace}                 
                  
 \def\PD      {\ensuremath{\mathrm{D}}\xspace}

 \def\PJ      {\ensuremath{\mathrm{J}}\xspace}                 
 \def\PK      {\ensuremath{\mathrm{K}}\xspace}

 \def\Pb      {\ensuremath{\mathrm{b}}\xspace}

 \def\Pe      {\ensuremath{\mathrm{e}}\xspace}

 \def\Pi      {\ensuremath{\mathrm{i}}\xspace}

 \def\Pp      {\ensuremath{\mathrm{p}}\xspace}

 \def\Ps      {\ensuremath{\mathrm{s}}\xspace}

 \def\thebaroffset{0.0em}
}
{
 
 \def\Pgamma      {\ensuremath{\gamma}\xspace}

 \def\Pmu         {\ensuremath{\mu}\xspace}

 \def\Ppi         {\ensuremath{\pi}\xspace}

 \def\Ppsi        {\ensuremath{\psi}\xspace}                 
                  
 \mathchardef\PDelta="7101
 \mathchardef\PXi="7104
 \mathchardef\PLambda="7103
 \mathchardef\PSigma="7106
 \mathchardef\POmega="710A
 \mathchardef\PUpsilon="7107
                  
 \def\PB      {\ensuremath{B}\xspace}                 
                  
 \def\PD      {\ensuremath{D}\xspace}

 \def\PJ      {\ensuremath{J}\xspace}                 
 \def\PK      {\ensuremath{K}\xspace}

 \def\Pb      {\ensuremath{b}\xspace}

 \def\Pe      {\ensuremath{e}\xspace}

 \def\Pi      {\ensuremath{i}\xspace}

 \def\Pp      {\ensuremath{p}\xspace}

 \def\Ps      {\ensuremath{s}\xspace}

 \def\thebaroffset{0.18em}
}
\newcommand{\offsetoverline}[2][\thebaroffset]{\kern #1\overline{\kern -#1 #2}}%

\makeatletter
\ifcase \@ptsize \relax
  \newcommand{\miniscule}{\@setfontsize\miniscule{4}{5}}
\or
  \newcommand{\miniscule}{\@setfontsize\miniscule{5}{6}}
\or
  \newcommand{\miniscule}{\@setfontsize\miniscule{5}{6}}
\fi
\makeatother

\DeclareRobustCommand{\optbar}[1]{\shortstack{{\miniscule (\rule[.5ex]{1.25em}{.18mm})}
  \\ [-.7ex] $#1$}}

\def\en         {{\ensuremath{\Pe^-}}\xspace}   
\def\ep         {{\ensuremath{\Pe^+}}\xspace}

\def\epem       {{\ensuremath{\Pe^+\Pe^-}}\xspace}

\def\mup        {{\ensuremath{\Pmu^+}}\xspace}
\def\mun        {{\ensuremath{\Pmu^-}}\xspace} 

\def\ellell     {\ensuremath{\ell^+ \ell^-}\xspace}

\def\g      {{\ensuremath{\Pgamma}}\xspace}

\def\squark    {{\ensuremath{\Ps}}\xspace}

\def\bquark    {{\ensuremath{\Pb}}\xspace}

\def\pion   {{\ensuremath{\Ppi}}\xspace}
\def\piz    {{\ensuremath{\pion^0}}\xspace}
\def\pip    {{\ensuremath{\pion^+}}\xspace}
\def\pim    {{\ensuremath{\pion^-}}\xspace}

\def\kaon    {{\ensuremath{\PK}}\xspace}
\def\Kbar    {{\ensuremath{\offsetoverline{\PK}}}\xspace}

\def\KorKbar {\kern \thebaroffset\optbar{\kern -\thebaroffset \PK}{}\xspace}

\def\Kp      {{\ensuremath{\kaon^+}}\xspace}
\def\Km      {{\ensuremath{\kaon^-}}\xspace}

\def\KS      {{\ensuremath{\kaon^0_{\mathrm{S}}}}\xspace}

\def\Kstarz  {{\ensuremath{\kaon^{*0}}}\xspace}
\def\Kstarzb {{\ensuremath{\Kbar{}^{*0}}}\xspace}

\def\D       {{\ensuremath{\PD}}\xspace}

\def\DorDbar {\kern \thebaroffset\optbar{\kern -\thebaroffset \PD}\xspace}

\def\Dp      {{\ensuremath{\D^+}}\xspace}
\def\Dm      {{\ensuremath{\D^-}}\xspace}

\def\DpDm    {\ensuremath{\Dp {\kern -0.16em \Dm}}\xspace}

\def\B       {{\ensuremath{\PB}}\xspace}
\def\Bbar    {{\ensuremath{\offsetoverline{\PB}}}\xspace}

\def\BorBbar {\kern \thebaroffset\optbar{\kern -\thebaroffset \PB}\xspace}
\def\Bz      {{\ensuremath{\B^0}}\xspace}
\def\Bzb     {{\ensuremath{\Bbar{}^0}}\xspace}
\def\Bd      {{\ensuremath{\B^0}}\xspace}
\def\Bdb     {{\ensuremath{\Bbar{}^0}}\xspace}
\def\BdorBdbar {\kern \thebaroffset\optbar{\kern -\thebaroffset \Bd}\xspace}
\def\Bu      {{\ensuremath{\B^+}}\xspace}

\def\Bs      {{\ensuremath{\B^0_\squark}}\xspace}

\def\BsorBsbar {\kern \thebaroffset\optbar{\kern -\thebaroffset \Bs}\xspace}

\def\jpsi     {{\ensuremath{{\PJ\mskip -3mu/\mskip -2mu\Ppsi}}}\xspace}

\def\Y#1S{\ensuremath{\PUpsilon{(#1S)}}\xspace}

\def\Lz          {{\ensuremath{\PLambda}}\xspace}

\def\LorLbar     {\kern \thebaroffset\optbar{\kern -\thebaroffset \PLambda}\xspace}

\def\Lb           {{\ensuremath{\Lz^0_\bquark}}\xspace}

\newcommand{\decay}[2]{\ensuremath{#1\!\to #2}\xspace} 
\def\ra                 {\ensuremath{\rightarrow}\xspace}
\def\to                 {\ensuremath{\rightarrow}\xspace}

\def\qsq       {{\ensuremath{q^2}}\xspace}

\def\CP                {{\ensuremath{C\!P}}\xspace}

\def\BdToKstmm    {\decay{\Bd}{\Kstarz\mup\mun}}
\def\BdToKstll    {\decay{\Bd}{\Kstarz\ellell}}
\def\BuToKll      {\decay{\Bu}{\Kp\ellell}}

\def\BdKstGam     {\decay{\Bd}{\Kstarz \g}}

\def\BdKstee  {\decay{\Bd}{\Kstarz\epem}}

\def\bsll     {\decay{\bquark}{\squark \ell^+ \ell^-}}
\def\AFB      {\ensuremath{A_{\mathrm{FB}}}\xspace}
\def\FL       {\ensuremath{F_{\mathrm{L}}}\xspace}
\def\AT#1     {\ensuremath{A_{\mathrm{T}}^{#1}}\xspace}           

\def\ctl       {\ensuremath{\cos{\theta_\ell}}\xspace}
\def\ctk       {\ensuremath{\cos{\theta_K}}\xspace}

\def\C#1      {\ensuremath{\mathcal{C}_{#1}}\xspace}                       
\def\Cp#1     {\ensuremath{\mathcal{C}_{#1}^{'}}\xspace}                    
\def\Ceff#1   {\ensuremath{\mathcal{C}_{#1}^{\mathrm{(eff)}}}\xspace}        
\def\Cpeff#1  {\ensuremath{\mathcal{C}_{#1}^{'\mathrm{(eff)}}}\xspace}       
\def\Ope#1    {\ensuremath{\mathcal{O}_{#1}}\xspace}                       
\def\Opep#1   {\ensuremath{\mathcal{O}_{#1}^{'}}\xspace}                    

\newcommand{\aunit}[1]{\ensuremath{\text{\,#1}}}       

\newcommand{\tev}{\aunit{Te\kern -0.1em V}\xspace}
\newcommand{\gev}{\aunit{Ge\kern -0.1em V}\xspace}
\newcommand{\mev}{\aunit{Me\kern -0.1em V}\xspace}
\newcommand{\kev}{\aunit{ke\kern -0.1em V}\xspace}
\newcommand{\ev}{\aunit{e\kern -0.1em V}\xspace}
\newcommand{\gevgev}{\ensuremath{\gev^2}\xspace} 
\newcommand{\mevc}{\ensuremath{\aunit{Me\kern -0.1em V\!/}c}\xspace}
\newcommand{\gevc}{\ensuremath{\aunit{Ge\kern -0.1em V\!/}c}\xspace}
\newcommand{\mevcc}{\ensuremath{\aunit{Me\kern -0.1em V\!/}c^2}\xspace}
\newcommand{\gevcc}{\ensuremath{\aunit{Ge\kern -0.1em V\!/}c^2}\xspace}

\def\fb   {\ensuremath{\aunit{fb}}\xspace}
\def\invfb   {\ensuremath{\fb^{-1}}\xspace}

\def\deriv {\ensuremath{\mathrm{d}}}

\def\gsim{{~\raise.15em\hbox{$>$}\kern-.85em
          \lower.35em\hbox{$\sim$}~}\xspace}
\def\lsim{{~\raise.15em\hbox{$<$}\kern-.85em
          \lower.35em\hbox{$\sim$}~}\xspace}

\def\pt         {\ensuremath{p_{\mathrm{T}}}\xspace}

\def\rad{\aunit{rad}}

\def\geant      {\mbox{\textsc{Geant4}}\xspace}

\def\tell1  {TELL1\xspace}
\def\ukl1   {UKL1\xspace}



%% file: title-LHCb-PAPER.tex

\begin{titlepage}
\pagenumbering{roman}

\vspace*{-1.5cm}
\centerline{\large EUROPEAN ORGANIZATION FOR NUCLEAR RESEARCH (CERN)}
\vspace*{1.5cm}
\noindent
\begin{tabular*}{\linewidth}{lc@{\extracolsep{\fill}}r@{\extracolsep{0pt}}}
\ifthenelse{\boolean{pdflatex}}
{\vspace*{-1.5cm}\mbox{\!\!\!\includegraphics[width=.14\textwidth]{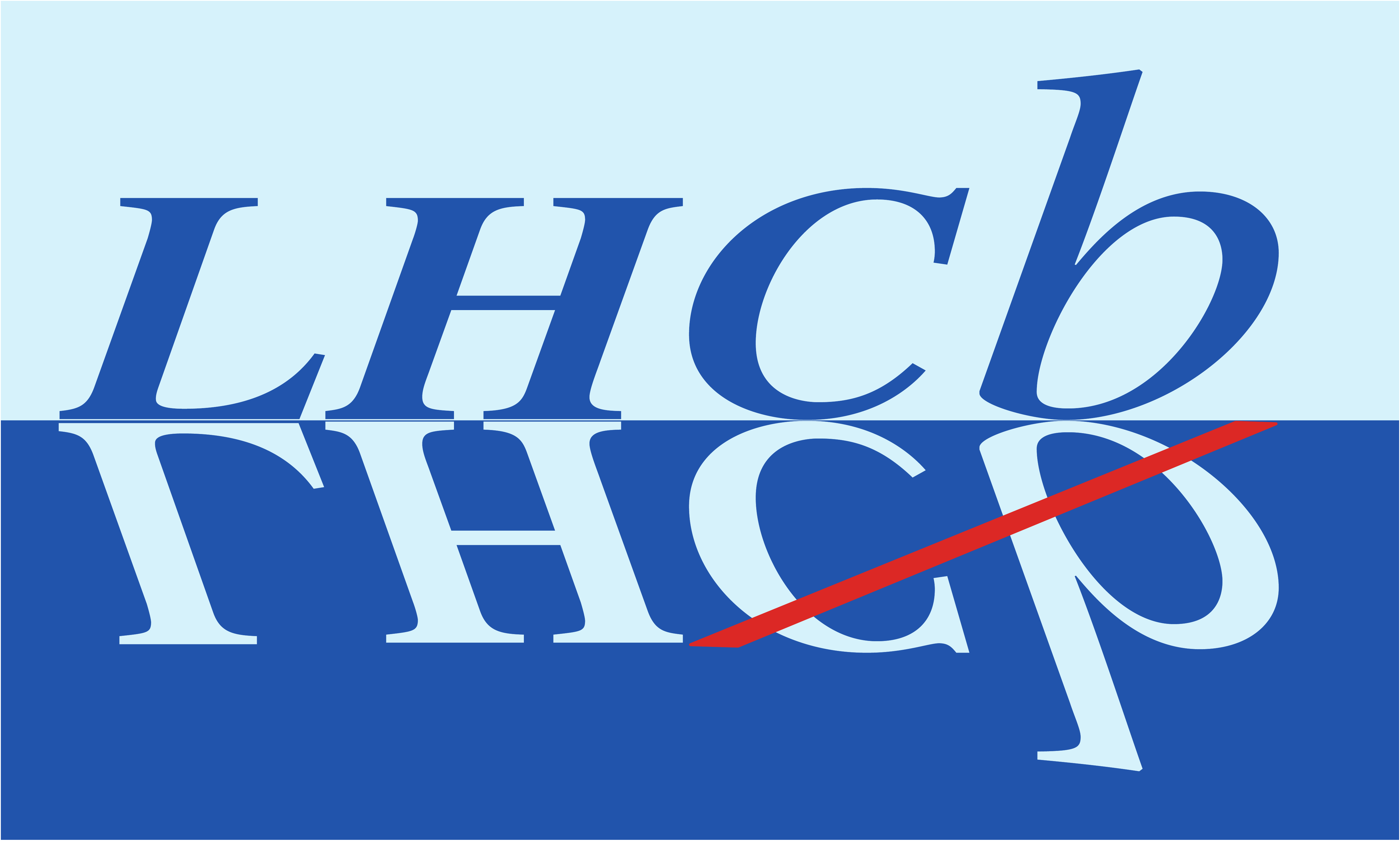}} & &}%
{\vspace*{-1.2cm}\mbox{\!\!\!\includegraphics[width=.12\textwidth]{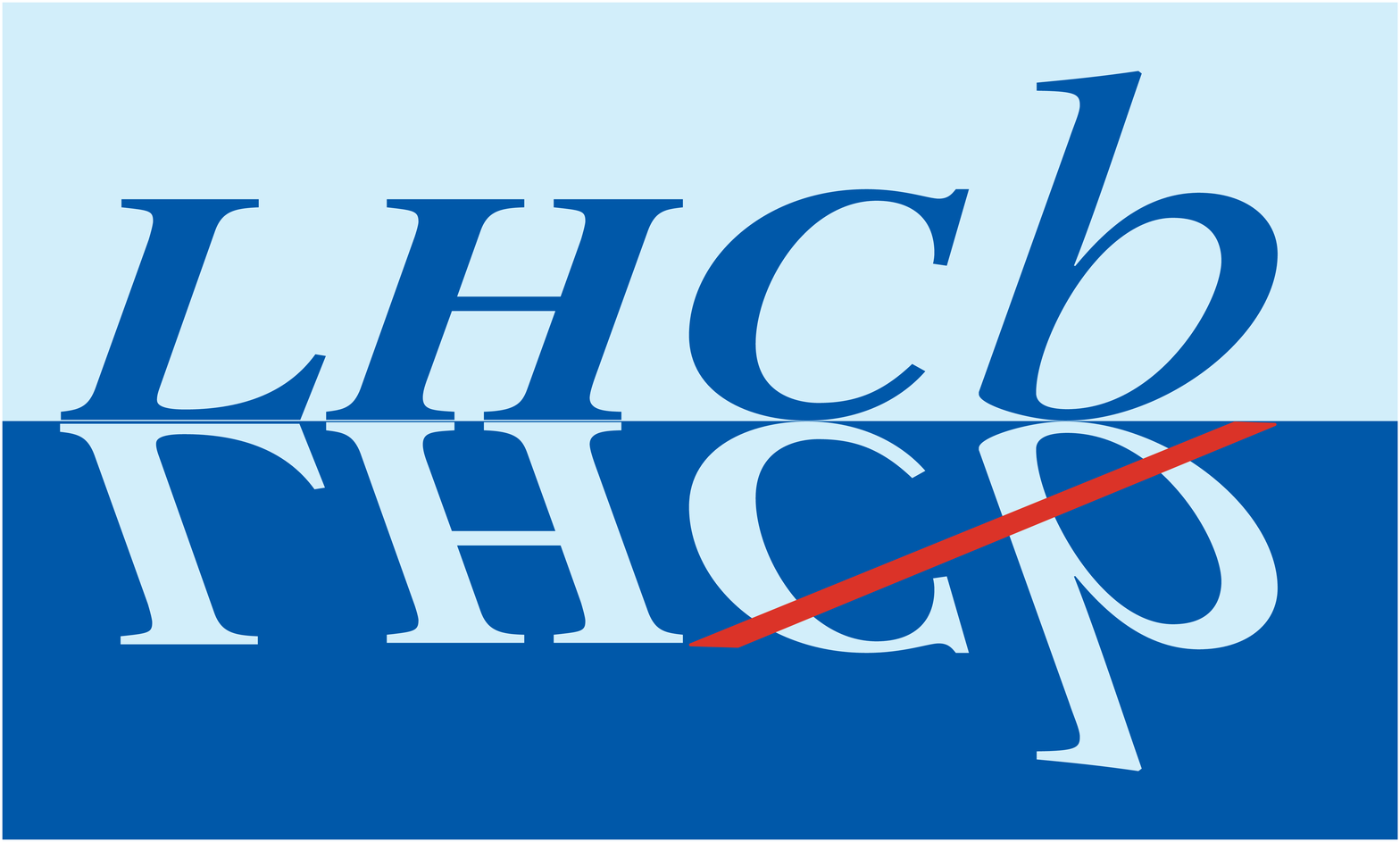}} & &}%
\\
 & & CERN-EP-2020-176 \\  
 & & LHCb-PAPER-2020-020 \\  
 & & 12 October 2020 \\ 
 & & \\
\end{tabular*}

\vspace*{3.0cm}

{\normalfont\bfseries\boldmath\huge
\begin{center}
  \papertitle
\end{center}
}

\vspace*{1.5cm}

\begin{center}
\paperauthors\footnote{Authors are listed at the end of this paper.}
\end{center}

\vspace{\fill}

\begin{abstract}
  \noindent
  An angular analysis of the $B^0 \to K^{*0} e^+ e^-$ decay is performed using a data sample corresponding to an integrated luminosity of 9${\mbox{\,fb}^{-1}}$ of $pp$ collisions collected with the LHCb experiment. The analysis is conducted in the very low dielectron mass squared ($q^2$) interval between $0.0008$ and $0.257\gevgev$, where the rate is dominated by the $B^0\to K^{\ast 0}\gamma$ transition with a virtual photon. The fraction of longitudinal polarisation of the $K^{\ast 0}$  meson, $F_{\rm L}$, is measured to be $F_{\rm L} = (4.4 \pm 2.6 \pm 1.4)\%$, where the first uncertainty is statistical and the second systematic. The $A_{\rm T}^{\rm Re}$ observable, which is related to the lepton forward-backward asymmetry, is measured to be $\ATRe=-0.06 \pm 0.08 \pm 0.02$.
  The $A_{\rm T}^{(2)}$ and $A_{\rm T}^{\rm Im}$ transverse asymmetries, which are sensitive to the virtual photon polarisation, are found to be $A_{\rm T}^{(2)} = 0.11 \pm 0.10 \ \pm 0.02$ and $A_{\rm T}^{\rm Im} = 0.02 \pm 0.10 \pm 0.01$. The results are consistent with Standard Model predictions and provide the world's best constraint on the $b\to s\gamma$ photon polarisation.
\end{abstract}

\vspace*{2.0cm}

\begin{center}
  Published in JHEP \textbf{12} (2020) 081
\end{center}

\vspace{\fill}

{\footnotesize
\centerline{\copyright~\papercopyright. \href{\paperlicenceurl}{\paperlicence}.}}
\vspace*{2mm}

\end{titlepage}


\newpage
\setcounter{page}{2}
\mbox{~}
%
%
%
%

%% file: body.tex
\section{Introduction}
\label{sec:Introduction}

Decay processes mediated by $b\to s\gamma$ transitions are suppressed in the Standard Model (SM) as they proceed through flavour-changing neutral currents involving electroweak-loop Feynman diagrams. The precise study of their properties is sensitive to small contributions from physics beyond the Standard Model (BSM).
Since in the SM the weak force only couples to left-handed quarks, the photons emitted in $b\to s\gamma$ transitions are predominantly left-handed. The contribution with right-handed polarisation is suppressed by the ratio of the $s$ and $b$ quark masses. Therefore, a larger right-handed contribution would represent a clear sign of BSM physics~\cite{Atwood:1997zr, Everett:2001yy, Grinstein:2004uu, Foster:2006ze, Lunghi:2006hc,Goto:2007ee, Becirevic:2012dx, Kou:2013gna, Haba:2015gwa, Paul:2016urs}.
The chirality of the $b\to s\gamma$ transition was indirectly probed at the \babar, \belle and \lhcb experiments, using measurements of the inclusive $B\to X_s\gamma$ branching ratio~\cite{Aubert:2007my, Lees:2012ym, Lees:2012wg, Saito:2014das, Belle:2016ufb} as well as the mixing-induced \CP asymmetries and time-dependent decay rates of radiative \Bd and \Bs decays~\cite{Ushiroda:2006fi, Aubert:2008gy, LHCb-PAPER-2019-015}. In this paper, the $b\to s\ell\ell$ transition where the dilepton pair originates from a virtual photon is used to measure the $b\to s\gamma$ photon polarisation. In order to isolate \bsll transitions dominated by the $b\to s\gamma$ contribution, the analysis is restricted to a region of very low dilepton mass squared (\qsq), which can only be accessed via the $b\to s\epem$ transition due to the low electron mass~\cite{Grossman:2000rk,Jager:2012uw}. This paper presents an angular analysis of the \BdToeeKst decay\footnote{The inclusion of charge-conjugate processes is implied throughout this paper. Natural units with $c = 1$ are used throughout this paper.} in the region of \qsq between 0.0008 and 0.257\gevgev. The symbol \Kstarz denotes the $\kaon^{*0}(892)$ meson reconstructed via its decay $\Kstarz\to\Kp\pim$.
An angular analysis of \BdToeeKst decays was performed by LHCb in the \qsq region between 0.002 and 1.120\gevgev~\cite{LHCb-PAPER-2014-066}.
The analysis presented here uses a data sample collected between the years 2011 and 2018. This sample comprises approximately five times as many $B^0$ decays. This analysis also employs a selection technique that greatly improves the signal purity as well as the sensitivity to the $b\to s\gamma$ photon polarisation.

The \BdToeeKst decay can be described over the full \qsq range by the left (right)-handed Wilson coefficients ${C}_7^{(\prime)}$, ${ C}_9^{(\prime)}$ and ${ C}_{10}^{(\prime)}$~\cite{Straub:2015ica, Ball:2006eu}. These coefficients encode information about short-distance effects and are sensitive to BSM physics.
 The detailed description of the \BdToKstll differential decay rate involves hadronic form factors describing the $\Bz\to\Kstarz$ transition and other long-distance effects that can be difficult to predict~\cite{Jager:2014rwa, Lyon:2014hpa, Ciuchini:2015qxb, Bobeth:2017vxj}. The results of the angular analysis of the \BdToKstmm decay ~\cite{LHCb-PAPER-2020-002}, the measurements of the branching fractions of several \bsll decays~\cite{LHCb-PAPER-2014-006,LHCb-PAPER-2015-023,LHCb-PAPER-2015-009} and the ratio of the branching fractions of the electron and muon channels of \BdToKstll and \BuToKll decays ~\cite{LHCb-PAPER-2019-009,LHCb-PAPER-2017-013} exhibit tensions with respect to SM predictions. Model independent fits of the \bsll measurements involve all of the Wilson coefficients mentioned above~\cite{Alguero:2019ptt, Aebischer:2019mlg, Crivellin:2018yvo, Arbey:2019duh, Ciuchini:2019usw}. Since the very-low-\qsq region is associated with the left- and right-handed electromagnetic operators~\cite{Grossman:2000rk}, it contains unique information that can be used to determine the $C_7$ and $C_7^\prime$ Wilson coefficients.

For the $\Bd\to\Kstarz\epem$ decay the partial decay width can be described in terms of \qsq and three angles $\theta_{\ell},\theta_{K}$ and $\phi$.  The angle $\theta_{\ell}$ is defined as the angle between the direction of the \ep and the direction opposite to that of the \Bz meson in the dielectron rest frame. The angle $\theta_{K}$ is defined as the angle between the direction of the kaon and the direction opposite to that of the \Bz meson in the \Kstarz meson rest frame. The angle $\phi$ is the angle between the plane containing the electron and positron and the plane containing the kaon and pion in the \Bz meson rest frame.  The basis is chosen so that the angular definition for the \Bzb decay is the \CP conjugate of that of the \Bz decay. These definitions are identical to those used for the \BdToKstmm analysis in Ref.~\cite{LHCb-PAPER-2013-019}, including the sign flip of $\phi$ ($\phi \to -\phi$) for the \Bdb decay. The angle $\phi$ is transformed such that $\phit=\phi+\pi$ if $\phi<0$. This transformation cancels out terms that have a $\sin\phi$ or $\cos\phi$ dependence and simplifies the angular expression without any loss of sensitivity to the remaining observables. In the region of \qsq considered in this paper, where the photon is almost on-shell, the fraction of $\Kp\pim$ pairs in an S-wave configuration is suppressed with respect to the value measured at higher \qsq~\cite{LHCb-PAPER-2016-012}, because a longitudinally polarised $\Kp\pim$ pair cannot couple to a real photon. Using Refs.~\cite{Kruger:2005ep, Lu:2011jm} it can be shown that the ratio of the S-wave fraction to the fraction of longitudinal polarisation of the $\Kstarz$ is constant as function of \qsq in the 0-6\gevgev range. Neglecting the $\Kp\pim$ S-wave contribution, and in the limit of massless leptons (a very good approximation for electrons), the \BdToeeKst angular distribution can be expressed as
\begin{equation}
  \label{eq:AngDistr}
  \begin{split}
  \frac{1}{\deriv(\Gamma+\bar\Gamma)/\deriv q^2} \frac{\deriv^4(\Gamma+\bar\Gamma)}{\deriv q^2\,\deriv{\cos\theta_\ell}\,\deriv{\ctk}\,\deriv{\phit}} =
\frac{9}{16\pi} \Big[
  & \tfrac{3}{4}(1-\FL)\sin^2\theta_K+\FL\cos^2\theta_K \\
  \phantom{\Big[} +&  \tfrac{1}{4}(1-\FL)\sin^2\theta_K\cos2\theta_\ell - \FL\cos^2\theta_K \cos2\theta_\ell\\
    \phantom{\Big[} +& (1-\FL)\ATRe\sin^2\theta_K \ctl\\
  \phantom{\Big[} +& \tfrac{1}{2}(1-\FL)\ATD\sin^2\theta_K \sin^2\theta_\ell\cos 2\phit \\
  \phantom{\Big[} +& \tfrac{1}{2}(1-\FL) \ATIm\sin^2\theta_K \sin^2\theta_\ell\sin 2\phit \Big]\,.
\end{split}
\end{equation}

\noindent The four angular observables \FL, \ATRe, \ATD and \ATIm are  combinations of the transversity amplitudes $A_0$, $A_\perp$ and $A_{||}$, as detailed in Ref.~\cite{LHCb-PAPER-2014-066}.
The observable $\FL$ corresponds to the longitudinal polarisation fraction of the \Kstarz meson and is expected to be small at low \qsq, since the virtual photon is quasi-real and therefore transversely polarised. The observable \ATRe is related to the lepton forward-backward asymmetry, \AFB, by $\ATRe = \frac{4}{3}\AFB(1-\FL)$~\cite{Becirevic:2011bp}.
The observable $\ATD$ is averaged between \Bz and \Bzb decays, while, given the $\phi$ sign flip for \Bdb decays, $\ATIm$ corresponds to a \CP asymmetry~\cite{Bobeth:2008ij}. The \ATRe, \ATD and \ATIm transverse asymmetries are related to the $P_i$ angular basis~\cite{DescotesGenon:2012zf} through $\ATRe= 2P_2$, $\ATD= P_1$ and $\ATIm= -2P_3^{\CP}$.

The $\ATD$ and $\ATIm$ observables depend only on the \BdKstee transversity amplitudes, $A_\perp$ and $A_{||}$, and vanish if these amplitudes are completely left-handed. In the limit $\qsq \to 0$, which is a good approximation for the \qsq region considered in this paper, the $\ATD$ and $\ATIm$ observables are closely related to the photon polarisation in  \BdKstGam transitions. In particular, the ratio of the right- and left-handed photon amplitudes, $A_{\rm R}$ and $A_{\rm L}$, can be related to \ATD and \ATIm through~\cite{Becirevic:2011bp, Paul:2016urs}
\begin{equation}
  \label{eq:polarisation}
  \begin{split}
     \tan\chi &\equiv \left| A_{\rm R}/A_{\rm L} \right|, \\
     \ATD &= \sin(2\chi)\cos(\phi_{\rm L}-\phi_{\rm R}), \\
     \ATIm &= \sin(2\chi)\sin(\phi_{\rm L}-\phi_{\rm R}),
  \end{split}
\end{equation}
where $\phi_{\rm L(R)}$ is the $A_{\rm L(R)}$ weak phase and the small strong phase difference between the amplitudes is neglected. Corrections to these approximations, due to terms proportional to ${C}_9$ and ${ C}_{10}$, are smaller than 0.006 even in the presence of large BSM effects in ${C}_7^{(\prime)}$~\cite{Straub:2018kue}. The mixing-induced \CP asymmetries and time-dependent decay rates of radiative \Bd and \Bs decays have very similar expressions, but also involve $B$-mixing phases~\cite{Paul:2016urs}.

 The $\ATD$ and $\ATIm$ observables are predicted to be very small in the SM when compared to the current experimental sensitivity.  Using the \flavio software package~\cite{Straub:2018kue} (version 2.0.0) the following SM predictions are calculated for the four angular observables in the \qsq range considered
\begin{equation}
  \label{eq:predictions}
  \begin{split}
     \FL({\rm SM})    &= \phantom{+}0.051\pm 0.013, \\
     \ATRe({\rm SM})  &= -0.0001 \pm 0.0004,\\
     \ATD({\rm SM})   &= \phantom{+}0.033\pm 0.020, \\
     \ATIm ({\rm SM}) &= -0.00012\pm 0.00034.
  \end{split}
\end{equation}
A detailed discussion of the theoretical uncertainties on the hadronic contributions involved in these predictions can be found in Ref.~\cite{Paul:2016urs}. The uncertainties on the transverse asymmetries are much smaller than the experimental sensitivity of the results presented in this paper.

\section{The \lhcb detector and data set} \label{Sec:Detector}
The study reported here is based on $pp$ collision data, corresponding to an integrated luminosity of 9\invfb, collected at the Large Hadron Collider (LHC) with the \lhcb detector~\cite{LHCb-DP-2008-001}. The data were taken in the years 2011, 2012 and 2015--2018, at centre-of-mass energies of 7, 8 and
13 TeV, respectively.
The \lhcb  detector~\cite{LHCb-DP-2008-001,LHCb-DP-2014-002} is a single-arm forward spectrometer covering the \mbox{pseudorapidity} range $2<\eta <5$. The detector includes a tracking system consisting of a vertex detector surrounding the $pp$ interaction region and of tracking stations on either sides of a $4{\rm\,Tm}$ dipole magnet. Charged particles are identified using information from two ring-imaging Cherenkov detectors (\rich),  electromagnetic (\ecal) and hadronic (HCAL) calorimeters and muon chambers. The online event selection is performed by a trigger, which consists of a hardware stage, based on information from the calorimeters and muon system, followed by a software stage, which fully reconstructs the event.
The hardware electron trigger requires the presence of an \ecal cluster with minimum transverse energy between 2.5 and 3.0\gev, depending on the data-taking period. Signal \BdKstee candidates are retained if at least one of the electrons fires the electron trigger. Alternatively, candidates are selected if the hardware trigger requirements were passed by objects in the rest of the event that are independent of the decay products of the signal \Bz candidate.
The software trigger requires a two-, three- or four-track vertex with a significant displacement from a primary $pp$ interaction vertex (PV). At least one charged particle must have a significant transverse momentum (\pt) and be inconsistent with originating from any PV. A multivariate algorithm~\cite{BBDT} is used to identify displaced vertices consistent with the decay of a \bquark hadron.

Samples of simulated events, produced with the software described in Refs.~\cite{Sjostrand:2006za,*Sjostrand:2007gs, LHCb-PROC-2010-056, Lange:2001uf, Golonka:2005pn, Allison:2006ve, *Agostinelli:2002hh, LHCb-PROC-2011-006}, are used to characterise signal and background contributions.
The simulated samples are corrected for known differences between data and simulation in kinematics, particle identification, detector occupancy, hardware trigger efficiency and reconstruction effects, based on a general approach developed by the $\lhcb$ collaboration for tests of lepton universality~\cite{LHCb-PAPER-2017-013}.

\section{Reconstruction and selection}
\label{Sec:Selection}
The \BdToeeKst candidates are formed by combining a \Kstarz candidate with a pair of oppositely charged tracks identified as electrons. For events passing the trigger, \Kstarz candidates are formed by combining  a pair of charged tracks identified as \Kp and \pim mesons. Each track is required to be of good quality and to be inconsistent with originating from a PV. Kaons and pions are required to have transverse momenta larger than 250\mev and are identified using information from the \rich detectors. Electrons with \pt exceeding 500\mev and with a good-quality vertex are used to form dielectron candidates.
The reconstructed invariant mass of the $\Kp\pim$ system is required to be within $100 \mev$ of the mass of the $\Kstarz$ meson~\cite{PDG2020}.

The tracks from the electrons, kaon and pion are required to form a good-quality vertex that is significantly displaced from any PV. In events with multiple PVs, the one with the smallest value of $\chi^2_{\rm IP}$ is associated to the \BdToeeKst candidate. Here $\chi^2_{\rm IP}$ is defined as the difference in the vertex-fit $\chi^2$ of a given PV
reconstructed with and without the tracks forming the candidate under consideration. In addition, the angle between the \Bd-candidate momentum vector and the vector between the associated PV and the \Bd decay vertex is required to be small.

Electrons can lose a significant amount of their energy when interacting with the detector material due to emission of bremsstrahlung photons. A dedicated procedure, which searches for neutral energy deposits in the \ecal that are compatible with being emitted by the electron upstream of the magnet, is applied to correct for this effect~\cite{LHCb-PAPER-2017-013}. The limitations of this recovery technique degrade the resolution of the reconstructed invariant masses of both the dielectron pair and the \Bd candidate.

\section{Analysis strategy} \label{sec:Strategy}
The \qsq region under study is chosen to maximise the sensitivity to $b\to s\gamma$ contributions (${C}_7^{(\prime)}$).
First of all, the reconstructed  invariant-mass squared resolution of the dielectron pair is improved by a kinematic fit that constrains the $\Kp\pim\ep\en$ mass to the known \Bd mass~\cite{PDG2020}.
The reconstructed \qsq is required to be lower than $0.25\gevgev$ to minimise the sensitivity to vector and axial-vector currents (${C}_9^{(\prime)}$ and ${C}_{10}^{(\prime)}$) while retaining as many signal candidates as possible. The larger data set of this analysis makes it possible to significantly reduce this upper bound with respect to Ref.~\cite{LHCb-PAPER-2014-066}, where it was 1\gevgev.
Low-\qsq signal candidates are most sensitive to ${C}_7^{(\prime)}$, but suffer from a degradation of the  resolution in $\phit$ due to multiple scattering of the quasi-collinear electrons in the tracking detectors. Furthermore, \BdKstGam decays followed by a photon conversion in the material of the detector contaminate the lower end of the $q^2$ spectrum. The reconstructed \qsq is thus required to exceed $10^{-4}\gevgev$, resulting in a \phit resolution of $0.11\rad$ and a \BdKstGam background fraction of about $2\%$ (see Sec.~\ref{Sec:Backgrounds}).

\begin{figure}[tbp]
  \centering
  \includegraphics[angle=0,width=0.55\textwidth]{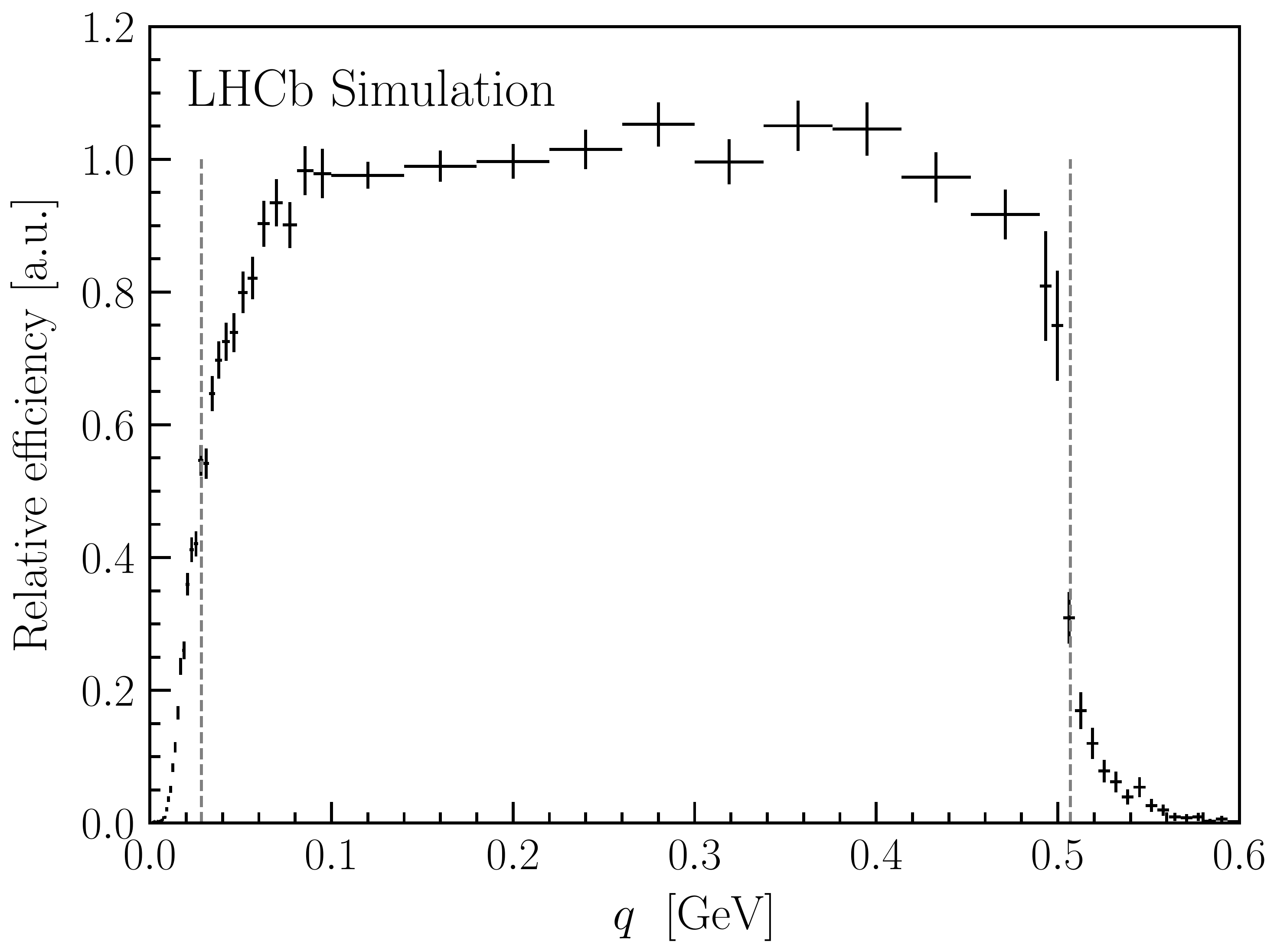}
 \caption{\small Relative efficiency as a function of the dielectron invariant mass ($q$). The points represent the efficiency obtained from simulation, while the vertical lines represent the effective \qsq boundary defined in the text.}
 \label{fig:effectiveq2}
\end{figure}

The signal selection efficiency as a function of the dielectron mass, obtained from simulation, is presented in Fig.~\ref{fig:effectiveq2}. The efficiency is approximately uniform across the signal region. Close to the boundaries, the efficiency drops due to the selected range of reconstructed \qsq and the effect of the dielectron mass resolution. Therefore, following Ref.~\cite{LHCb-PAPER-2014-066}, effective \qsq boundaries are defined between 0.0008\gevgev and 0.257\gevgev to allow for theoretical predictions of the angular observables without input from \lhcb simulation. Using the \flavio software package, it was checked that predictions with both SM and BSM values for the Wilson coefficients calculated in this effective \qsq range (grey line in Fig.~\ref{fig:effectiveq2}) agree very well with those calculated taking into account a complete description of the \qsq efficiency using \lhcb simulation (points in Fig.~\ref{fig:effectiveq2}).

The region of reconstructed $q^2$ below $10^{-4}\gevgev$ is enriched in \BdKstGam decays and is used as a control sample. Its kinematics and background level are very similar to the signal \qsq region, but with much larger candidate yields.

The \BdToeeKst branching ratio is expected to be as small as $(2.0\pm 0.4)\times 10^{-7}$ in the \qsq range studied in this paper~\cite{Straub:2018kue}. Nonetheless, a pure signal sample is obtained by greatly reducing all expected background contributions with the selection described in Sec.~\ref{Sec:Backgrounds}. A fit to the reconstructed $\Kstarz\epem$ invariant mass, \meeKstar, in a wide range between 4500 and 6200\mev is used to estimate the remaining background contamination, as explained in Sec.~\ref{sec:MassFit}.
Afterwards, a four-dimensional fit to the $\Kp\pim\epem$ invariant mass and the three angles \ctl, \ctk and \phit is used to measure the four angular observables \FL, \ATRe, \ATD and \ATIm. This fit is performed in a reduced \meeKstar\ window between 5000 and 5400\mev in order to simplify the angular modelling of the background components. The background fractions are constrained to those obtained in the wider \meeKstar\ window. Signal and background angular shapes are determined using simulation and data samples (Sec.~\ref{sec:angModel}) and then fitted to the signal sample (Sec.~\ref{sec:fitRes}).

The \meeKstar\ mass resolution, the angular acceptance and the background rates depend on how the event has been triggered at the hardware level. The data sample is therefore divided into two categories: events for which at least one of the two electron candidates fires the electron trigger, and events triggered by activity in the event that is not associated with any of the signal decay particles. Furthermore, the data sets collected in 2011--12 (Run 1) and 2015--18 (Run 2) are treated separately to account for kinematic differences due to the different $pp$ collision energies.

\section{Background studies}
\label{Sec:Backgrounds}
Several sources of specific background are considered, with studies performed using samples of simulated events unless stated otherwise. All of the identified background sources that are expected to contribute at a level of more than $1 \%$ of the signal are modelled and included in the analysis.

A large background comes from the semileptonic \decay{\Bd}{\Dm\ep\nu} decay, with  ${\Dm \ra \Kstarz \en { \overline \nu}}$. This contribution populates the region below the \Bd mass and has a combined branching fraction four orders of magnitude larger than that of the signal. In the case where both neutrinos have low energies, the signal selection is ineffective at rejecting it. The positron from the \Bd decay tends to be more energetic than the electron from the \Dm decay and hence the reconstructed \ctl distribution favours large values since \ctl is highly correlated with the \epem energy asymmetry. In order to avoid any potential bias in the measurement of \ATRe, a symmetric requirement of $|\ctl|<0.8$ is applied, resulting in a $5\%$ loss of signal while rejecting $98 \%$ of this semileptonic background.

The radiative decay \BdKstGam has a branching fraction about two orders of magnitude larger than that of the signal and has a very similar distribution in the reconstructed $\Kp\pim\ep\en$ mass. In the signal sample, contamination from this background is at the level of $23 \%$, but is reduced to about $2 \%$ by rejecting dielectron pairs compatible with originating from detector material~\cite{LHCb-DP-2018-002}.
A specific weighting procedure is applied to the $\Bd \ra \Kstarz \g$ simulation to match the true \epem mass distribution of Ref.~\cite{Tsai:1973py} since the \geant version used here does not accurately model high-mass \epem pair production.

The $\Bd \ra \Kstarz \eta$ and $\Bd \ra \Kstarz \piz$ decays where the $\eta$ or $\piz$ meson decays to $\ep \en \g$ (Dalitz decay) can pass the selection if the photon is very soft, or if it is recovered as a bremsstrahlung photon. In the latter case, the \meeKstar\ mass peaks at the \Bd mass.  The contamination from the $\eta$ ($\piz$) Dalitz decay is estimated to be at the level of 4\% (2\%) in the mass region used in the angular fit.

To suppress background from \BsToPhiee decays, where the $\phi$ meson decays to a $\Kp\Km$ pair and one of the kaons is misidentified as a pion, the two-hadron invariant mass computed under the $K^+K^-$ hypothesis is required to be larger than 1040\mev.
Background contributions from misidentified \LbToeeL, $\Bd \ra \Kstarz \pip \pim$ and \decay{\Bdb}{\Kstarzb\epem} decays are found to be negligible.

Partially reconstructed (PR) background contributions arising from $\decay{B}{\Kstarz \pi \epem}$ decays, where one of the pions is not reconstructed, are suppressed by exploiting the kinematic balance of the decay. The ratio of the \Kstarz and dielectron momenta components transverse to the \Bd direction is expected to be unity unless some energy is lost through bremsstrahlung emission. Since at low \qsq bremsstrahlung photons do not significantly modify the dielectron direction, this ratio can be used to recover the lost energy and recompute the corrected reconstructed \Bd mass called $m_{\rm HOP}$~\cite{LHCb-PAPER-2017-013}. In the case of PR background, however, the missing particles are not necessarily emitted in the same direction as either electron. Therefore the requirement $m_{\rm HOP} > 4900 \mev$ rejects about $70 \%$ of the PR background with a $90 \%$ signal efficiency, estimated from simulation. This background yield is found to be 5\% of that of the signal in the narrow mass window.

In order to reduce the level of combinatorial background, a multivariate classifier based on a boosted decision tree algorithm \mbox{(BDT)~\cite{Breiman,AdaBoost}} is used. The BDT classifier is trained to separate simulated \BdKstee events from background events taken from the upper invariant-mass sideband ($\meeKstar > 5600\mev$) in data and uses eight kinematic and decay topology variables including the $\chi^2_{\rm IP}$ of final-state particles with respect to the associated PV and the $\pt$ of the \Bd candidate and its flight distance from the PV. The classifier achieves a background rejection of $90 \%$ and a signal efficiency of $94 \%$. The semileptonic and combinatorial background sources contribute a contamination of about 7\% and are therefore modelled in the fit to the data.

\section{{\boldmath $\Kp\pim\epem$} invariant-mass spectra}
\label{sec:MassFit}
An unbinned maximum-likelihood fit to the $\Kp\pim\epem$ invariant-mass distribution is performed simultaneously to the signal and control samples in order to measure the background fractions.
 In both samples, the \BdKstee and \BdKstGam components are described by a bifurcated Crystal Ball (CB) function~\cite{Skwarnicki:1986xj}, which consists of a Gaussian core with asymmetric power-law tails. The shapes of the $\Kstarz\eta$ and $\Kstarz\piz$ background contributions are modelled by non-parametric probability density functions (PDFs)~\cite{Cranmer:2000du}, while the shape of the PR background is modelled by the sum of a CB function and a Gaussian function. Finally, the shapes of the semileptonic and combinatorial background (SL/C) are parametrised together by an exponential function. All shapes apart from the SL/C are fixed from simulation. The widths and mean values of the signal CB functions are corrected for differences between data and simulation  using a high-purity data sample of \hbox{\BdToJPsieeKst} candidates.

\begin{figure}[p]
  \centering
  \includegraphics[angle=0,width=0.495\textwidth]{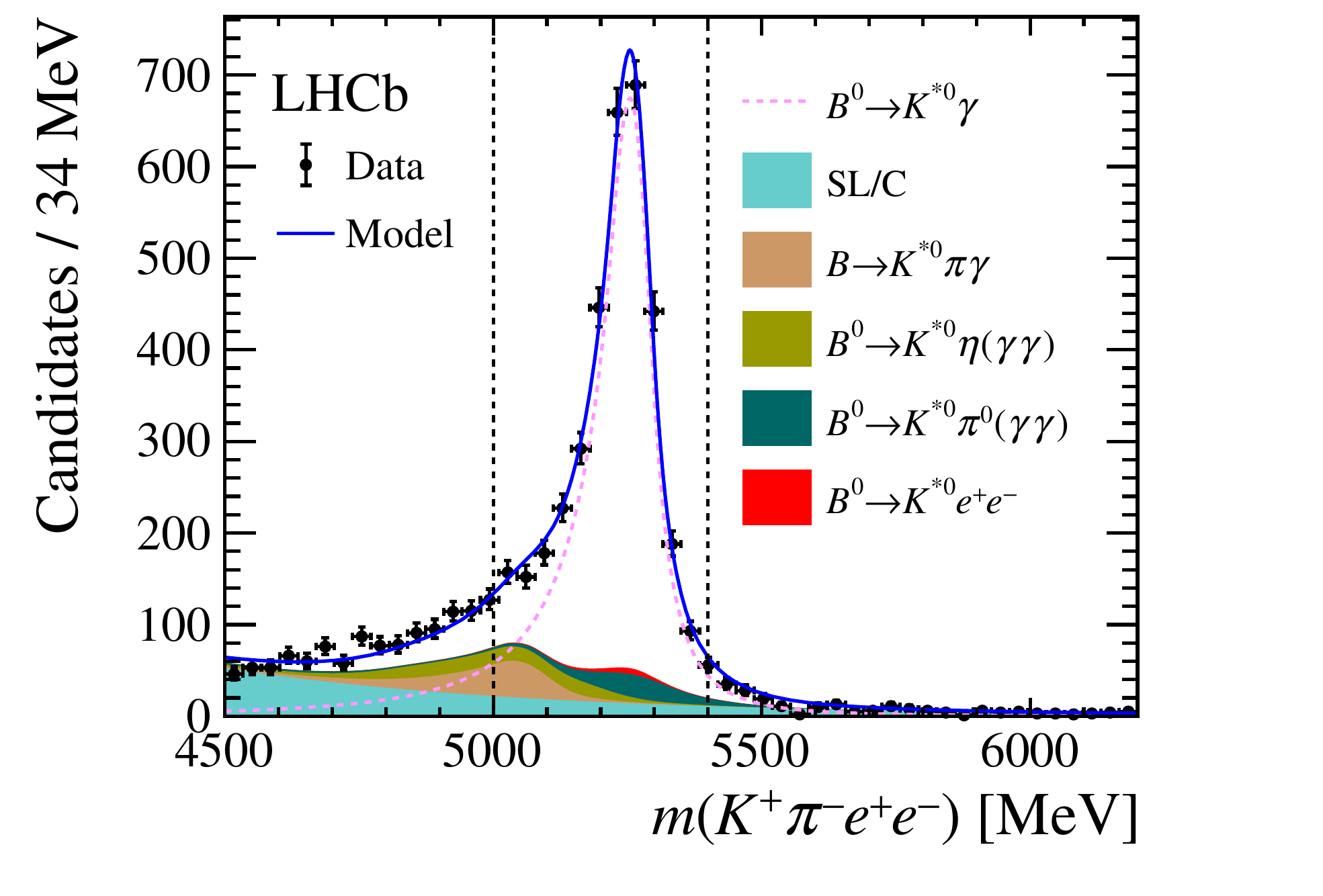}
  \includegraphics[angle=0,width=0.495\textwidth]{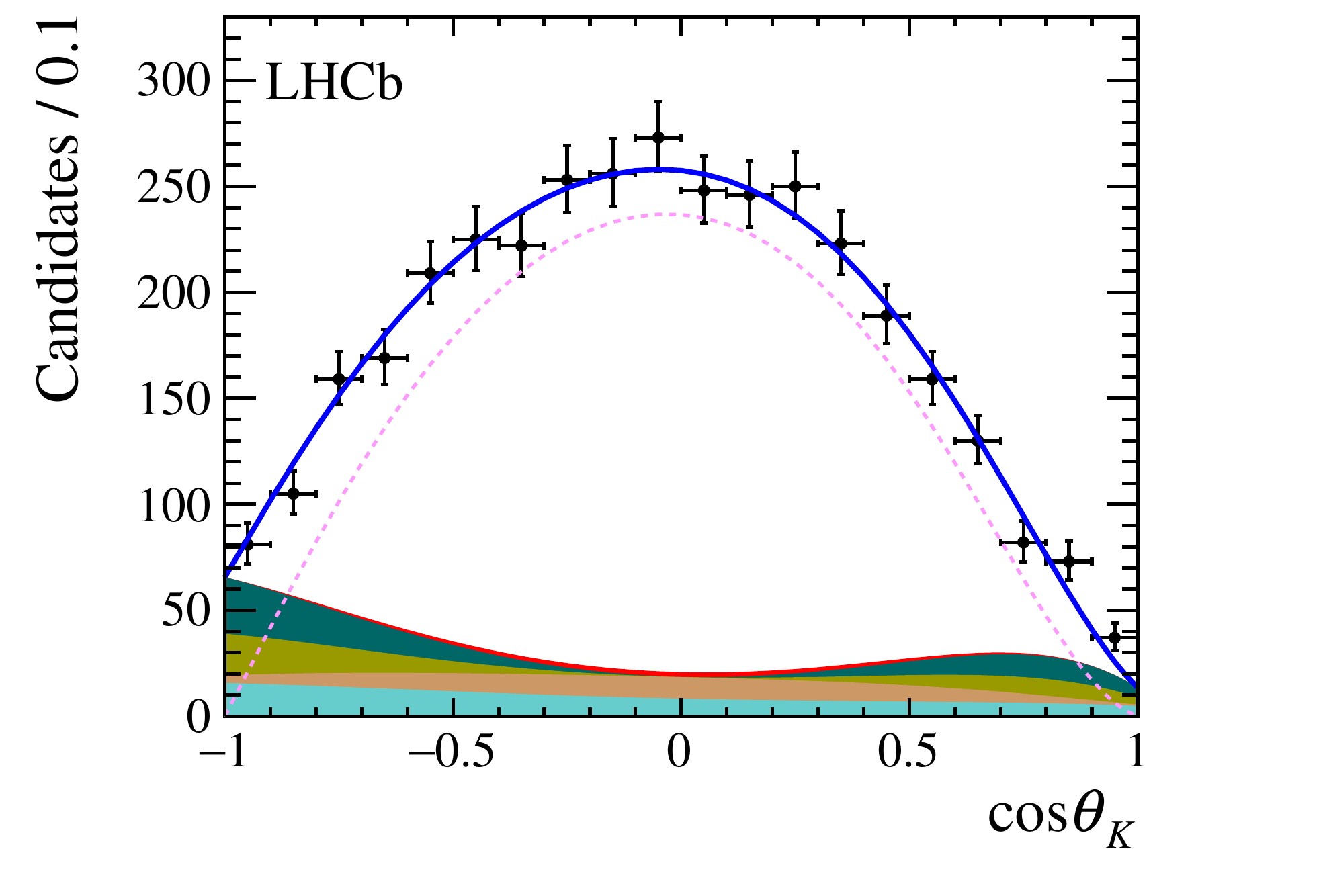}
 \caption{\small Distributions of the (left) $\Kp\pim\epem$ invariant mass and (right) \ctk of \BdKstGam candidates. The black points represent the data, while the solid blue curve shows the total PDF. The signal component is represented by the dashed pink line and the shaded areas are the background components, as detailed in the legend. The SL/C component is composed of semileptonic and combinatorial background contributions. The dashed vertical lines indicate the restricted mass range used in the angular analysis.}
 \label{fig:DataFitResultKstG}
\end{figure}

\begin{figure}[p]
  \centering
  \includegraphics[angle=0,width=0.495\textwidth]{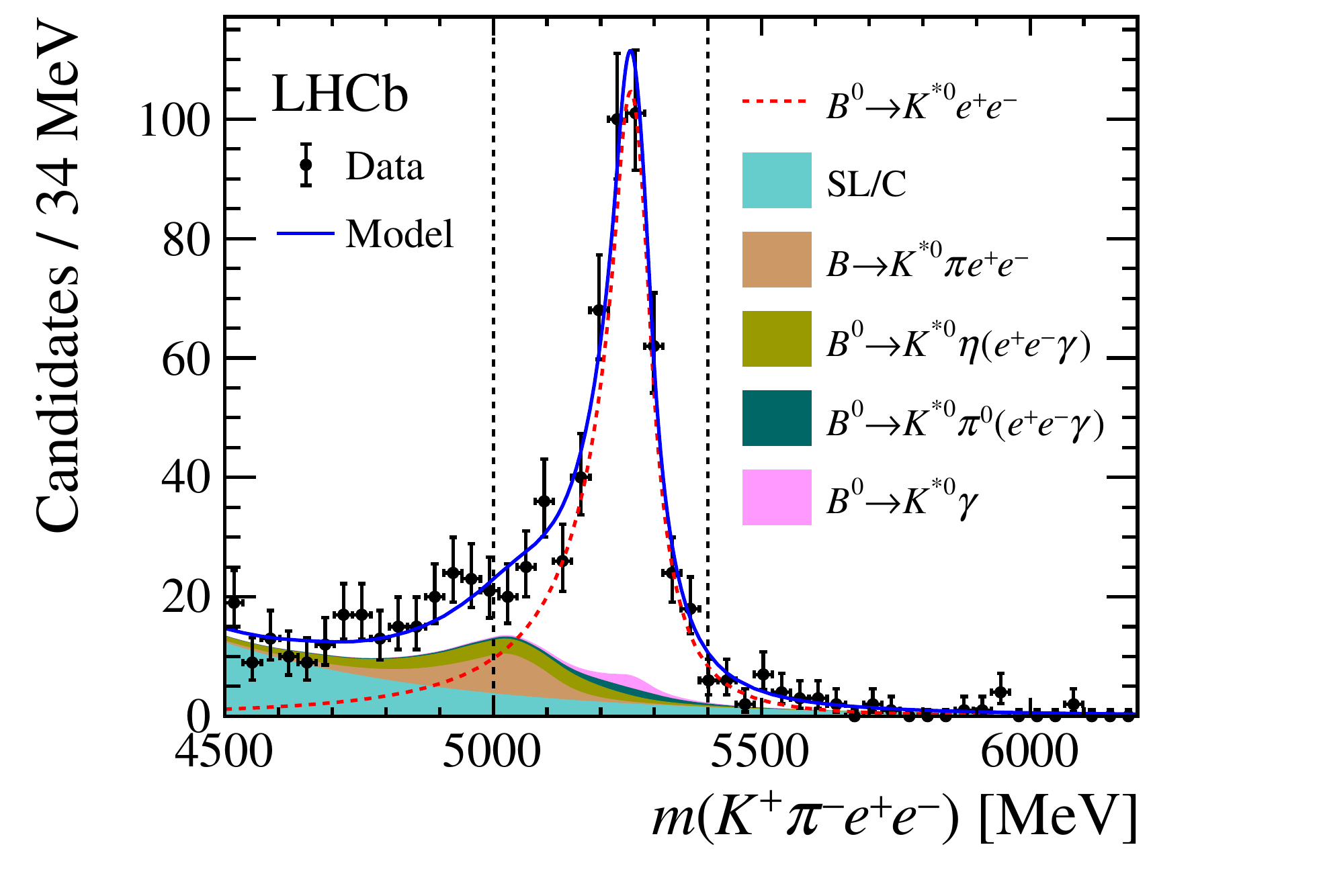}
  \includegraphics[angle=0,width=0.495\textwidth]{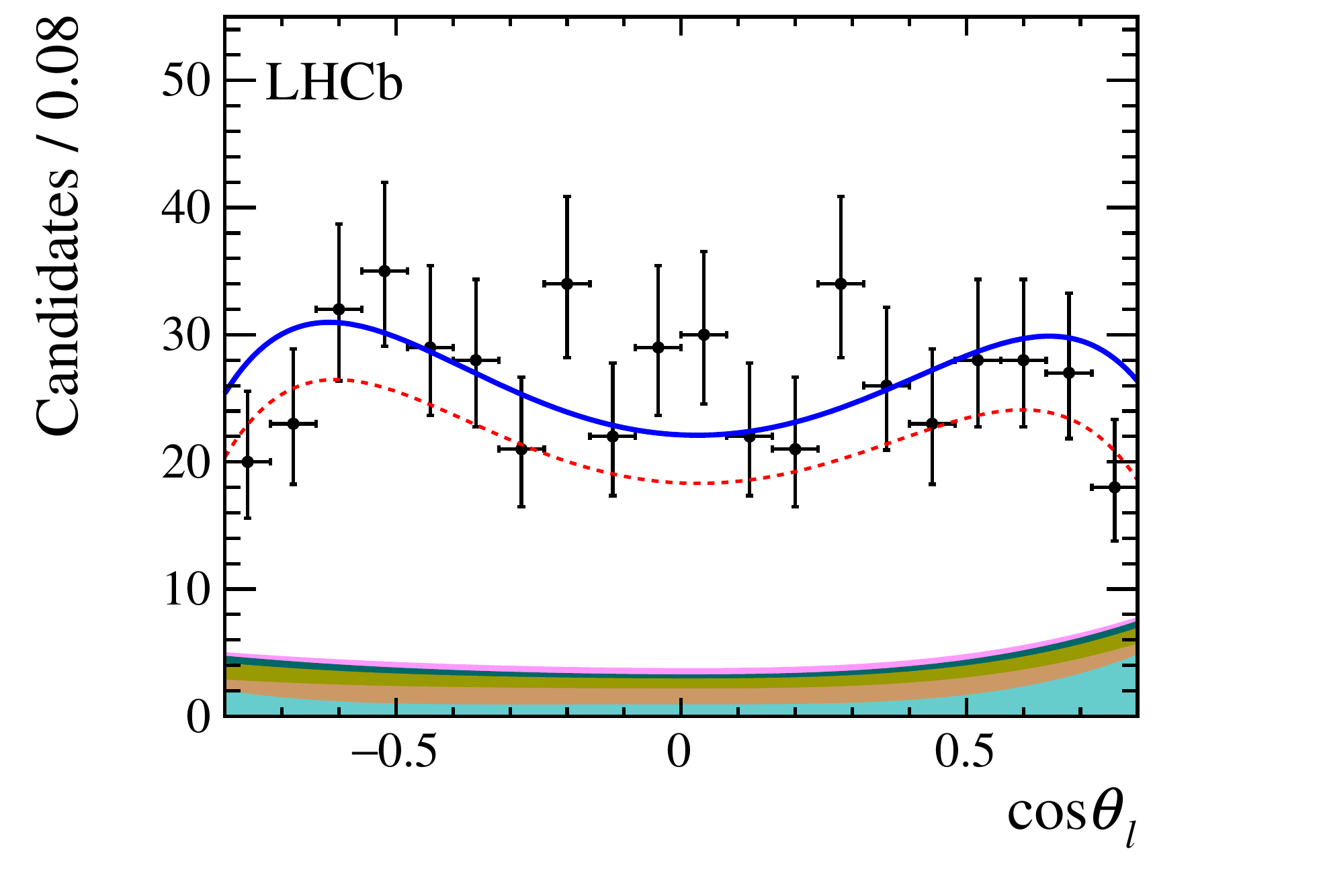}
  \includegraphics[angle=0,width=0.495\textwidth]{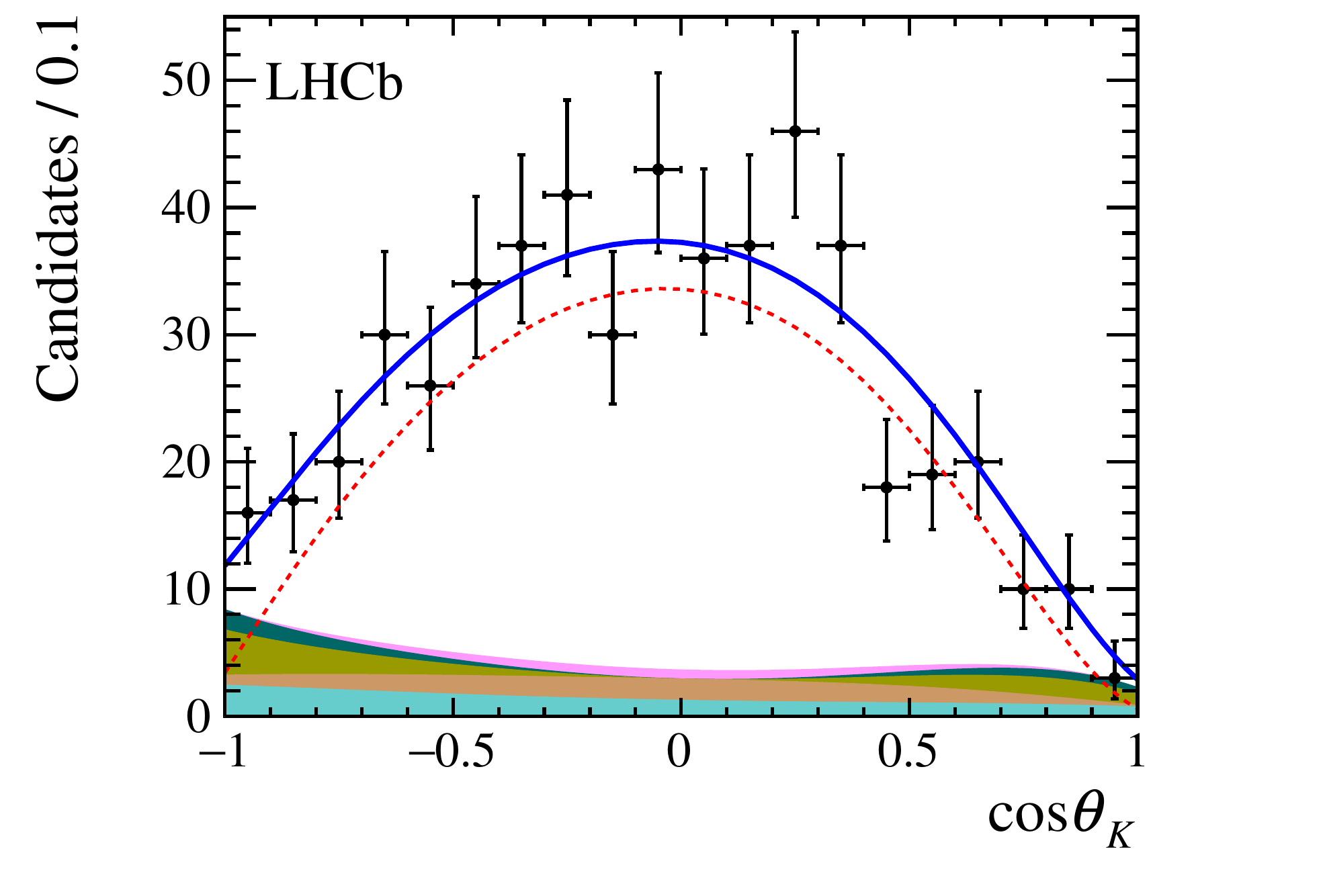}
  \includegraphics[angle=0,width=0.495\textwidth]{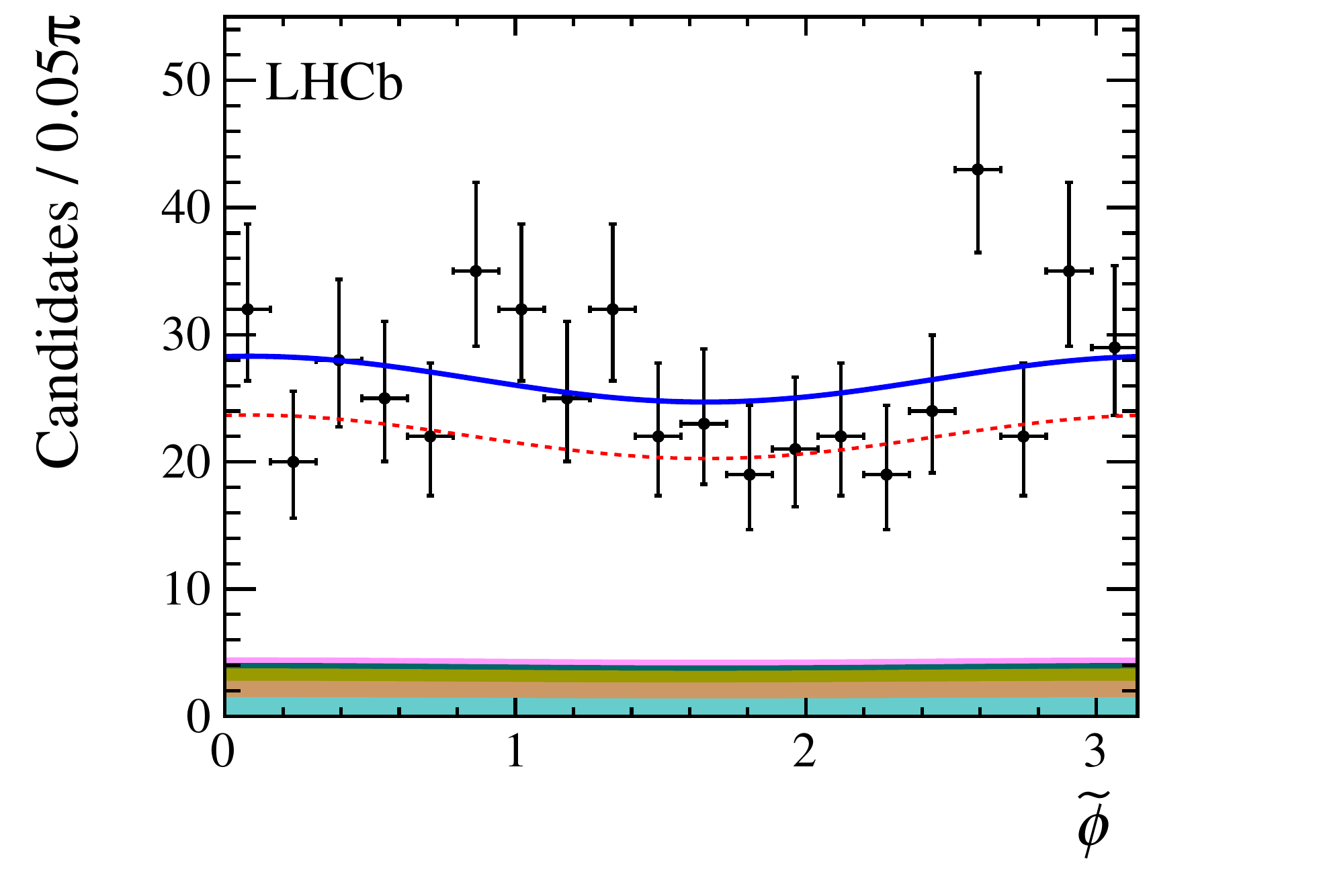}
 \caption{\small Distributions of the (top left) $\Kp\pim\epem$ invariant mass, (top right) \ctl, (bottom left) \ctk and (bottom right) \phit variables of \BdKstee candidates in the reconstructed \qsq range between $10^{-4}\gevgev$ and 0.25\gevgev. The black points represent the data, while the solid blue curve shows the total PDF. The signal component is represented by the dashed red line and the shaded areas are the background components, as detailed in the legend. The SL/C component is composed of semileptonic and combinatorial background contributions.  The dashed vertical lines indicate the restricted mass range used in the angular analysis.}
 \label{fig:DataFitResult}
\end{figure}

Since the $b\to s\gamma$ contribution dominates both the \BdKstee and $B\to\Kstarz\pi e^+e^-$ decay rates in the \qsq region considered, the PR background is expected to be similar for the signal and control samples. The ratio of PR background and signal yields is therefore shared between the two samples. Using the fit in the wider \meeKstar\ mass window, the \BdKstee yield in the restricted \meeKstar\ range used for the angular fit is estimated to be 450. In the control sample, the \BdKstGam yield is about 2950, while in both samples the signal-to-background ratio is about 5.
The invariant-mass distributions together with the PDFs resulting from the fit are shown in Figs.~\ref{fig:DataFitResultKstG} and~\ref{fig:DataFitResult} for the control and signal samples, respectively.

\section{Angular modelling}
\label{sec:angModel}
The \BdKstee angular distribution described using Eq.~\ref{eq:AngDistr} is multiplied by an acceptance function evaluated from simulated $\Bd\to\Kstarz\epem$ decays to take into account the effect of the reconstruction and selection efficiency. The acceptance function, $\varepsilon$, factorises between the angles such that
\begin{equation}
    \varepsilon (\ctl,\ctk,\phit) \simeq \varepsilon(\ctl)\varepsilon(\ctk)\varepsilon(\phit).
\end{equation}
The \ctk and \ctl acceptance functions are modelled with fourth-order Legendre polynomials. For the \phit angle, non-uniform acceptance terms proportional to $\cos(2\phit)$ and $\sin(2\phit)$ are allowed for completeness, however, no significant deviation from a uniform distribution is observed.

Since for \BdKstGam decays the presence of the electrons is only due to the interaction of the real photon with the detector material, the \ctl and \phit dependent parts of Eq.~\ref{eq:AngDistr} are integrated out to model purely the \ctk dependent part of the \BdKstGam angular shape, which depends only on the parameter \FL. The value of \FL for the \BdKstGam decay is obtained from the fit to the control sample detailed in Sec.~\ref{sec:fitRes}. When included as a background in the fit to the signal sample, the \BdKstGam angular shape is obtained from the simulation sample and is assumed to factorise between the angles.

The background contributions due to $\Bd \ra \Kstarz \eta$ and $\Bd \ra \Kstarz \piz$ decays that contribute to the signal and control samples are modelled using simulation and are also assumed to factorise in the three angles. Since this background has $\FL = 1$ due to the $\piz$ or $\eta$ angular momentum, the \ctk distribution has a very different shape compared to \BdKstee decays. Its precise modelling is validated by the measurement of the \FL parameter in the control channel reported in Sec.~\ref{sec:fitRes}.

The angular shape of the PR background is modelled using the same functional shape as the signal, determined from $\Bu \ra K_1(1270) \ep \en$ simulated events, where one of the pions from the $K_1(1270) \ra \Kp \pim \pip$ decay is not reconstructed.

A sample of $\Bd \ra \Kstarz \ep \mun$ candidates from \lhcb data is used to determine the SL/C angular shapes. Since this decay is forbidden in the SM due to lepton flavour conservation, this sample will mostly comprise semileptonic and combinatorial background events. The \qsq and BDT requirements are slightly relaxed to increase the sample size. It is checked that this sample is a good proxy for SL/C background by assigning the muon candidate an electron mass and comparing the resulting angular and $\Kp\pim\epem$ invariant-mass distributions to those of \BdKstee decays in the upper mass sideband and low BDT output regions.  The angular shapes of the three angles are found to factorise in this sample and therefore are modelled separately.

\section{Angular observables}
\label{sec:fitRes}
To determine the four angular observables, \FL, \ATD, \ATIm and \ATRe, an unbinned maximum likelihood fit is performed to the \meeKstar, \ctl, \ctk and \phit distributions in a restricted \meeKstar\ window between 5000 and 5400\mev. The inclusion of \meeKstar\ improves the statistical power of the fit since, even within the restricted window, the mass shapes of the signal and of the background contributions are very different. The fractions of the fit components are constrained using a multivariate Gaussian function to the results in the wide mass range extrapolated to the narrow mass range. Pseudoexperiments are used to assess the impact of fitting the \meeKstar\ distribution again in the restricted range. The resulting bias is found to be negligible.

The fitting procedure is verified using a large sample of fully simulated events, with the obtained values of \FL, \ATD, \ATIm and \ATRe in excellent agreement with the inputs.
The \BdToeeKst fit is then validated on data by performing a similar fit to the \meeKstar\ and \ctk distributions of the control sample. The \ctk distribution for the control sample, together with the PDF projections resulting from the fit, are shown in Fig.~\ref{fig:DataFitResultKstG}. The fitted value of $\FL = 0.0^{+0.7}_{-0.0}\,\%$ is compatible with a completely transverse \Kstarz polarisation, as expected due to presence of the real photon.
Finally, the \BdToeeKst fit is further validated using $10\,000$ pseudoexperiments including signal and background components. Several input values for the angular observables, \FL, \ATD, \ATIm and \ATRe, are studied including those associated with BSM models, and the results are in good agreement with the inputs, with the exception of \FL, where a small bias at the level of $7\%$ of its statistical uncertainty is observed and corrected for. Furthermore, the non-negligible size of the \phit resolution results in an underestimation of the magnitude of \ATD and \ATIm by $4\%$. Although this shift is negligibly small for the magnitudes expected in the SM, it could be sizeable for large ${C}_7^{\prime}$ values and therefore the \ATD and \ATIm values  are corrected for this effect.
The angular distributions, \ctl, \ctk and \phit, for the signal region, together with the PDF projections resulting from the fit, are shown in Fig.~\ref{fig:DataFitResult}.  Results for the angular observables are given in Sec.~\ref{sec:results}.

\section{Systematic uncertainties} \label{sec:systematics}

\begin{table}[b]
\renewcommand{\arraystretch}{1.2}
\centering
\caption{Summary of systematic uncertainties on the four angular observables, \ATD, \ATIm, \ATRe and \FL. The total systematic uncertainty is the sum in quadrature of all the contributions. For comparison, the statistical uncertainties are shown in the last row of the table.}
\label{tab:syst}
\begin{tabular}{l|cccc}
\hline
Source of systematic & \ATD & \ATIm & \ATRe & \FL \\
\hline
Simulation sample size for acceptance& $0.007$ & $0.007$ & $0.007$ & $0.003$\\
Acceptance function modelling & $0.004$ & $0.001$ & $0.008$ & $0.001$ \\
$\Bd \ra \Kstarz \ep \mun$ sample size for SL/C & $0.007$ & $0.007$ & $0.007$ & $0.003$ \\
SL/C angular modelling &  $0.012$ & $0.005$ & $0.006$ & $0.005$ \\
PR model other than $K_1(1270)$ & $0.001$  & $0.003$ & $0.002$ & $0.001$ \\
$\eta$ or $\piz$ angular modelling & $<0.001$ & $<0.001$ & $0.002$ & $0.010$ \\
Corrections to simulation & $0.003$ & $0.001$ & $0.003$ & $0.007$ \\
Signal mass shape & $0.002$ & $0.002$ & $0.004$ & $0.001$ \\
\hline
Total systematic uncertainty & $0.017$ & $0.012$ & $0.015$ & $0.014$ \\
Statistical uncertainty & $0.103$ & $0.102$ & $0.077$ & $0.026$ \\
\hline
\end{tabular} \\
\end{table}

To evaluate  systematic uncertainties resulting from the limited size of the data and simulation samples used to determine the angular shapes and acceptances, a bootstrapping technique is used~\cite{efron:1979}. In addition, systematic uncertainties related to various modelling choices used in the fits are evaluated by generating pseudoexperiments with an alternative model and fitting with the nominal model used to fit the data. The results of the mass and angular fits are then compared with the input values to assess the size of the uncertainties. The alternative modelling choices considered are detailed in the following.

The systematic uncertainties related to the corrections applied to simulated events used to model the angular acceptance are evaluated by fitting uncorrected simulated events. An alternative model using Legendre polynomials of order six instead of four is used to estimate the systematic uncertainties related to the shape of the acceptance function.

To take into account possible variations in the angular shapes of the PR background due to states other than the $K_1(1270)$ meson, alternative shapes are determined from a sample of $\Bu \ra \Kp \pim \pip \ep \en$ simulated events. This sample is reweighted in the $\Kp \pim \pip$ Dalitz plane to match the distribution in $\Bu \ra \jpsi (\ra \ep \en) K^{\rm res}(\ra \Kp \pim \pip)$ data, where $K^{\rm res}$ is any kaon resonance that decays to $\Kp\pim\pip$.
Alternative models for the SL/C background are obtained by tightening either the \qsq or BDT requirements used in the $\Bd \ra \Kstarz \ep \mun$ selection.

To estimate the systematic uncertainty due to differences between data and simulation in the \BdKstee mass shapes, the signal mass PDF is corrected using a fit to the \BdKstGam  rather than the \BdToJPsieeKst channel.
The  systematic uncertainties are summarised in Table~\ref{tab:syst}. The total systematic uncertainty, obtained by adding all individual sources in quadrature, is smaller than the statistical uncertainty for all observables.

\section{Results \label{sec:results}}
\begin{figure}[t]
  \centering
  \includegraphics[angle=0,width=0.85\textwidth]{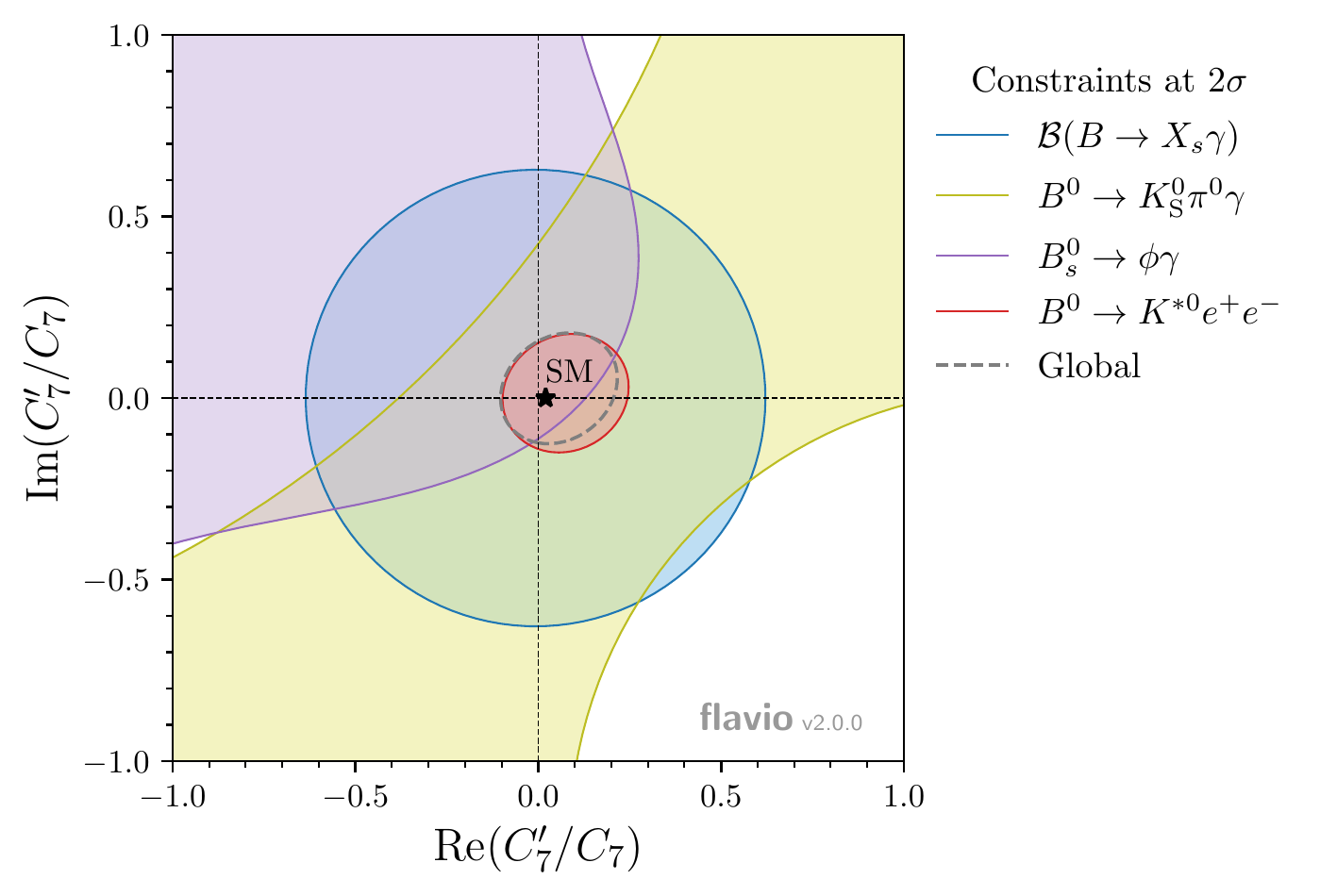}
 \caption{\small Constraints at $2 \sigma$ level on the real and imaginary parts of the ratio of right- and left-handed Wilson coefficients, $C_7^{\prime}$ and $C_7$. The $C_7$ coefficient is fixed to its SM value. The measurements of the inclusive branching fraction, ${\cal B}(B\to X_s\gamma)$, and the $\Bd\to\KS\piz\gamma$ mixing-induced \CP asymmetry by the \belle and \babar experiments~\cite{Aubert:2007my, Lees:2012ym, Lees:2012wg, Saito:2014das, Belle:2016ufb, Ushiroda:2006fi, Aubert:2008gy} are shown in blue and yellow, respectively, the $\Bs \ra \phi \g$ measurements at \lhcb
~\cite{LHCb-PAPER-2019-015} in purple and the measurement presented in this paper in red. The global fit is shown in  dashed lines and the SM prediction is represented by a black star and corresponds to the ratio of $s$- and $b$-quark masses.}
 \label{fig:flavio}
\end{figure}

The four \BdKstee angular observables measured in the effective \qsq range from $0.0008$ to $0.257$\gevgev are found to be
 \begin{eqnarray} \nonumber
  \FL &=&  0.044 \pm 0.026 \pm 0.014, \\\nonumber
  \ATRe &=& -0.06 \pm 0.08 \pm 0.02,  \\ \nonumber
  \ATD &=& +0.11 \pm 0.10 \pm 0.02, \\\nonumber
   \ATIm &=& +0.02 \pm 0.10 \pm 0.01,\nonumber
\end{eqnarray}
where the first contribution to the uncertainty is statistical and the second systematic. Correlations between the  observables are measured to be
\begin{center}
\begin{tabular}{c|rrrr}
      & $F_{\rm L}$ & \ATRe & \ATD & \ATIm \\ \hline
                   $F_{\rm L}$ &  1.00 & $-0.02$ & $-0.01$ &  0.02 \\
                   \ATRe &       &  1.00 &  0.05 &  0.02 \\
                   \ATD  &       &       &  1.00 &  0.10  \\
                   \ATIm &       &       &       &  1.00 \\
\end{tabular}
\end{center}
These results supersede Ref.~\cite{LHCb-PAPER-2014-066}. Using Eq.~\ref{eq:polarisation}, the measured \ATD and \ATIm observables are used to determine the photon polarisation in \BdKstGam decays
\begin{eqnarray} \nonumber
  {\rm Re}\left( A_{\rm R}/A_{\rm L}\right) &=& 0.05 \pm 0.05  \\ \nonumber
  {\rm Im}\left( A_{\rm R}/A_{\rm L}\right) &=& 0.01 \pm 0.05.   \nonumber
\end{eqnarray}
Furthermore, using the \flavio~software package~\cite{Straub:2018kue}, these measurements can be used to determine the polarisation of the $b\to s\gamma$ transition, which can be expressed as the ratio of the right- and left-handed $C_7^{(\prime)}$ Wilson coefficients. Details about the calculations of hadronic contributions can be found in Ref.~\cite{Paul:2016urs}.
The obtained constraints are shown in Fig.~\ref{fig:flavio}, where they are compared to those from previous measurements by the \belle, \babar and \lhcb experiments~\cite{Aubert:2007my, Lees:2012ym, Lees:2012wg, Saito:2014das, Belle:2016ufb, Ushiroda:2006fi, Aubert:2008gy, LHCb-PAPER-2019-015}. Here, the $C_7^{(\prime)}$ regularisation-scheme independent effective coefficients are calculated at the scale $\mu=4.8\gev$~\cite{Paul:2016urs}. The value of the left-handed $C_7$ coefficient is fixed to its SM value, $C_7^{\rm SM}=-0.2915$. Theoretical uncertainties related to the predictions of the experimental observables are taken into account in the constrained areas. The results presented in this paper provide the world’s best constraint on the $b\to s\gamma$ photon polarisation.

\section{Conclusion \label{sec:conclusion}}
An angular analysis of the \BdToeeKst decay is performed using proton-proton collision data, corresponding to an integrated luminosity of 9.0\invfb, collected by the \lhcb experiment between 2011 and 2018.
Angular observables are measured for the first time in the \qsq range from $0.0008$ to $0.257$\gevgev.

The results are consistent with SM predictions~\cite{Becirevic:2011bp, Jager:2014rwa, Straub:2018kue} and are used to measure both the real and imaginary parts of the \BdKstGam photon polarisation with a precision of 5\%.
Furthermore, the results of this paper make it possible to constrain the $b\to s\gamma$ photon polarisation with significantly better precision than the combination of previous measurements.

%% file: acknowledgements.tex
\section*{Acknowledgements}
%
%
\noindent We express our gratitude to our colleagues in the CERN
accelerator departments for the excellent performance of the LHC. We
thank the technical and administrative staff at the LHCb
institutes.
We acknowledge support from CERN and from the national agencies:
CAPES, CNPq, FAPERJ and FINEP (Brazil); 
MOST and NSFC (China); 
CNRS/IN2P3 (France); 
BMBF, DFG and MPG (Germany); 
INFN (Italy); 
NWO (Netherlands); 
MNiSW and NCN (Poland); 
MEN/IFA (Romania); 
MSHE (Russia); 
MICINN (Spain); 
SNSF and SER (Switzerland); 
NASU (Ukraine); 
STFC (United Kingdom); 
DOE NP and NSF (USA).
We acknowledge the computing resources that are provided by CERN, IN2P3
(France), KIT and DESY (Germany), INFN (Italy), SURF (Netherlands),
PIC (Spain), GridPP (United Kingdom), RRCKI and Yandex
LLC (Russia), CSCS (Switzerland), IFIN-HH (Romania), CBPF (Brazil),
PL-GRID (Poland) and OSC (USA).
We are indebted to the communities behind the multiple open-source
software packages on which we depend.
Individual groups or members have received support from
AvH Foundation (Germany);
EPLANET, Marie Sk\l{}odowska-Curie Actions and ERC (European Union);
A*MIDEX, ANR, Labex P2IO and OCEVU, and R\'{e}gion Auvergne-Rh\^{o}ne-Alpes (France);
Key Research Program of Frontier Sciences of CAS, CAS PIFI,
Thousand Talents Program, and Sci. \& Tech. Program of Guangzhou (China);
RFBR, RSF and Yandex LLC (Russia);
GVA, XuntaGal and GENCAT (Spain);
the Royal Society
and the Leverhulme Trust (United Kingdom).

%% file: LHCb_Authorship_07-Jul-2020.tex
\centerline
{\large\bf LHCb collaboration}
\begin
{flushleft}
\small
R.~Aaij$^{31}$,
C.~Abell{\'a}n~Beteta$^{49}$,
T.~Ackernley$^{59}$,
B.~Adeva$^{45}$,
M.~Adinolfi$^{53}$,
H.~Afsharnia$^{9}$,
C.A.~Aidala$^{84}$,
S.~Aiola$^{25}$,
Z.~Ajaltouni$^{9}$,
S.~Akar$^{64}$
% ...

{\footnotesize \it

% ...
$ ^{9}$Universit{\'e} Clermont Auvergne, CNRS/IN2P3, LPC, Clermont-Ferrand, France\\
% ...
$ ^{25}$INFN Sezione di Milano, Milano, Italy\\
% ...
$ ^{31}$Nikhef National Institute for Subatomic Physics, Amsterdam, Netherlands\\
% ...
$ ^{45}$Instituto Galego de F{\'\i}sica de Altas Enerx{\'\i}as (IGFAE), Universidade de Santiago de Compostela, Santiago de Compostela, Spain\\
% ...
$ ^{49}$Physik-Institut, Universit{\"a}t Z{\"u}rich, Z{\"u}rich, Switzerland\\
% ...
$ ^{53}$H.H. Wills Physics Laboratory, University of Bristol, Bristol, United Kingdom\\
% ...
$ ^{59}$Oliver Lodge Laboratory, University of Liverpool, Liverpool, United Kingdom\\
% ...
$ ^{64}$University of Cincinnati, Cincinnati, OH, United States\\
% ...
$ ^{67}$Syracuse University, Syracuse, NY, United States\\
% ...
$ ^{84}$University of Michigan, Ann Arbor, United States, associated to $^{67}$\\
\bigskip
% ...
\medskip
}
\end{flushleft}